\documentclass[fleqn,usenatbib]{mnras}

\usepackage{newtxtext,newtxmath}
\usepackage[T1]{fontenc}
\DeclareRobustCommand{\VAN}[3]{#2}
\let\VANthebibliography\thebibliography
\def\thebibliography{\DeclareRobustCommand{\VAN}[3]{##3}\VANthebibliography}

\usepackage{graphicx, subcaption}	
\usepackage{amsmath}	
\usepackage{soul}     
\usepackage[dvipsnames]{xcolor}
\usepackage{caption}
\usepackage{multicol}
\usepackage{relsize}  
\usepackage{tabularx, multirow}
\usepackage{fix-cm}
\usepackage{orcidlink}

\title[{High-[Mg/Fe] Milky Way bar stars}]{Dynamical properties of high-[Mg/Fe] stars in the Milky Way bar region}

\author[A. Pandey and O. Gerhard]{
Aakash Pandey,$^{1}$$^{,2}$\thanks{E-mail: apandey@mpe.mpg.de}\orcidlink{0009-0006-5677-7347}
Ortwin Gerhard,$^{1}$\thanks{E-mail: gerhard@mpe.mpg.de}\orcidlink{0000-0003-3333-0033}
\\
$^{1}$Max-Planck-Institute fur Extraterrestrische Physik, Gießenbachstraße, 85748 Garching, Germany
\\
$^{2}$Excellence Cluster ORIGINS, Boltzmannstr. 2, D-85748 Garching, Germany}

\date{Accepted XXX. Received YYY; in original form ZZZ}
\pubyear{2025}

\begin{document}
\label{firstpage}
\pagerange{\pageref{firstpage}--\pageref{lastpage}}
\maketitle

\begin{abstract}
The origin of the high-$\alpha$ component of the Galactic bulge remains debated, unlike the bar-driven origin of the low-$\alpha$ bulge. We examine the metallicity-dependent dynamical properties of high-[Mg/Fe] stars in the bar region, using samples of low- and high-[Mg/Fe] stars from APOGEE DR17, complemented by the PIGS catalogue of ${\rm [Fe/H]}<-1$ stars. The mean Galactocentric rotational velocity $\overline{V}_{\phi}(R)$ is nearly cylindrical for both low- and high-[Mg/Fe] stars across the bulge and outer bar. $\overline{V}_{\phi}(R)$ of high-[Mg/Fe] stars with ${\rm [Fe/H]}\geq-0.6$ is similar within errors to low-[Mg/Fe] stars in the bulge, and 10-20\% lower in the outer bar. The mean radial velocity field of these stars exhibits a quadrupole pattern similar to low-[Mg/Fe] stars. Integrating orbits in realistic barred Galactic potentials, these model-independent properties correspond to a peanut bulge in the orbital density distributions for high-[Mg/Fe] stars with ${\rm [Fe/H]}\geq-0.6$, transitioning toward a more spheroidal structure at lower metallicities. Additionally, $\overline{V}_{\phi}({\rm [Fe/H]})$ for stars increases steeply as metallicity increases from about [Fe/H] $\sim-1.3$, resembling the spin-up observed at larger Galactic radii. This is accompanied by a transition in the dominant orbit families, from co- and counter-rotating ${\rm cloud\,A}$ and ${\rm x_4}$ orbits at low metallicities to co-rotating bar-supporting ${\rm x_1}$ family tree, ${\rm box}$, and ${\rm cloud\,A}$ orbits at solar metallicity. Our results strengthen the case that the bulk of the high-[Mg/Fe] component in the bar region evolved from an $\alpha$-enhanced disc, while metal-poor stars with ${\rm [Fe/H]}<-1$ trace a more turbulent origin.
\end{abstract}


\section{Introduction}
The central region of the Milky Way is crucial for understanding its formation and evolutionary history. Large-scale surveys suggest that multiple structural components -- such as the bar, boxy/peanut-shaped bulge, thick and thin discs, halo, and recently identified proto-Galactic stars -- coexist within the inner $\sim5$ kpc, commonly referred to as the Galactic bar region.

Stars in the Galactic bar region exhibit a clear chemical bimodality in the [Fe/H]-[Mg/Fe] (or ${\rm [\alpha/Fe]}$) plane \citep{Gonzalenz2011Alpha, Rojas2019bimodal, Lian2020milky, Queiroz2021milky}.  This bimodality is characterized by two main populations: a metal-rich low-$\alpha$ component $\left({\rm [Fe/H]}\gtrsim-0.3,\,{\rm [Mg/Fe]\lesssim0.15}\right)$ and a metal-intermediate high-$\alpha$ component $\left({\rm -1\lesssim[Fe/H]}\lesssim0,\,{\rm [Mg/Fe]\gtrsim0.15}\right)$. Together, these populations dominate the Milky Way bulge. At lower metallicities ${\rm [Fe/H]}\lesssim-1$, stars are part of the inner stellar halo and the proto-Galactic population \citep{Belokurov2022Dawn, Rix2022poor, Chandra_2024, Arentsen2024orbital}.

\cite{Nataf2010Xshape} and \cite{McWilliam2010Xshape} found that the red clump stars near the minor axis of the Galactic bulge exhibit a bimodal magnitude distribution along the line of sight, most prominently for stars with ${\rm [Fe/H]}\ge0$ \citep{Ness2012origin}. Kinematically, metal-rich stars have colder kinematics and display cylindrical rotation \citep[][]{Howard2009, Ness2013kinematics, RojasArriagada2020many}. These observed properties can be reproduced in $N$-body simulations of a bulge formed through the secular evolution of a disc that developed a bar and subsequently a boxy/peanut bulge \citep{Shen2010, DiMatteo2015milky, Debattista2017fractionation}. Insights from such models, combined with the chemical similarity between the metal-rich low-$\alpha$ bulge stars and the thin disc in the solar neighbourhood, strongly support the idea that this bulge population originated from an early low-$\alpha$ thin disc via secular evolution \citep{Combes1990, Debattista2006, Shen2010}.

However, a longstanding debate persists over whether the high-$\alpha$ metal-intermediate stars of the Galactic bulge are an inward extension of the solar neighbourhood high-$\alpha$ thick disc. The key question underlying this discussion is whether the metal-intermediate, high-$\alpha$ bulge originated from the secular evolution of the thick disc or from a separate early intense star formation. 

\cite{Ness2012origin} found that red clump stars with $-0.5<{\rm [Fe/H]}<0$ also exhibit a bimodal magnitude distribution along the line of sight.  At the time, due to the lack of $\alpha$-abundance measurements, these stars were interpreted as a part of the early thin disc, redistributed into the boxy/peanut bulge by the bar instability \citep[][]{Ness2013Stellar}. However, this scenario was later challenged by the APOGEE data, which revealed a significant $\alpha$-enhancement in these stars \citep{Rojas2019bimodal, Queiroz2020bule2disc}. 

On the other hand, the location of the knee in the [$\alpha$/Fe]-[Fe/H] plane was found to be more metal-rich in the bulge than in the solar neighbourhood \citep[][]{CunhaSmith2006bulge, Rojas2017complex} -- implying a higher early star formation rate in the bulge than in the solar neighbourhood thick disc. However, later APOGEE data suggested that the knee’s position is nearly constant across both regions \citep{Zasowski2019Apogee, Imig2023tale}. Using APOGEE DR16, \cite{Queiroz2021milky} estimated the probability of bulge stars being on bar-supporting ${\rm x_1}$ orbits based on their orbital frequency ratios and found a significant fraction of high-$\alpha$ bulge stars on ${\rm x_1}$ orbits. However, they interpreted these high-$\alpha$ barred stars as belonging to a spheroidal bulge formed through early intense star formation, later dynamically captured by the Galactic bar.

Several lines of evidence indicate that high-$\alpha$ stars in the Galactic bulge are part of the Galactic bar and boxy/peanut bulge -- their cylindrical line-of-sight velocity pattern \citep{Ness2016ApogeeKinematics, RojasArriagada2020many, Wylie2021a2a}, the orbital characterization in \citet{Queiroz2021milky}, and the quadrupole pattern in the mean radial velocity field \citep{Liao2024insights}. However it may be possible that a pre-existing spheroidal bulge formed during an intense early star formation period and trapped later by the low-$\alpha$ bar could lead to similar observational characteristics \citep{Saha_2012, Saha_2013}. To investigate this further we will study the metallicity-dependent observational signatures based on full 6D phase-space information.

Lastly, metal-poor stars ([Fe/H]$\lesssim-1$) were not well studied in the bulge region until recently. Thanks to surveys like PIGS \citep{Arentsen2020pristine} and the Gaia XP spectra \citep{Carrasco2021xp, DeAngeli2022}, a larger sample of metal-poor stars down to [Fe/H] $\sim-3$ is now available. Using the PIGS data, \cite{Arentsen2024orbital} showed that a large fraction ($\sim70\%$) of these stars is confined to the bulge region \citep[see also][]{Rix2022poor}. Although these stars have dispersion-dominated kinematics, they still have significant rotation, with a mean velocity of $\sim80\,{\rm km\,s^{-1}}$ at [Fe/H] $\sim-1$ which decreases to $\sim40\,{\rm km\,s^{-1}}$ at [Fe/H] $\sim-2.2$. This study grouped these stars into two spheroidal components: a more metal-rich, faster-rotating group and a more metal-poor, slower-rotating group. However, the reason for the high rotation at these metallicities is still unclear.

In this work, we use APOGEE DR17 stars to study the dynamical properties of high-$\alpha$ stars in the Galactic bar region as a function of metallicity, and thereby investigate their evolutionary history. To extend the analysis to lower metallicities ([Fe/H] $\lesssim-1$), we include stars from the PIGS catalogue. Section~\ref{sec: Data} describes the data samples and the correction for the photometric selection in APOGEE. Section~\ref{sec: Results} presents the results, first examining the rotational properties and mean radial velocity fields of the observed stars, followed by an analysis of their orbital properties using realistic dynamical Made-to-Measure potential models. Finally, we discuss our results in section~\ref{sec: Discussion} and summarize them in section~\ref{sec: Summary}.

\section{Data}\label{sec: Data}
In this work, we use samples of stars from the APOGEE spectroscopic survey\footnote{Apache Point Obs. Galactic Evolution Experiment, \cite{Majewski2016apogee}}data release DR17 \citep{Abdurro2022apogeeDR17}, and from the PIGS spectroscopic survey\footnote{Pristine Inner Galaxy Survey, \protect{\cite{Arentsen2020pristine}}} of low-metallicity stars as released by \protect{\cite{Arentsen2024orbital}}. Chemical abundances and radial velocities provided by these surveys are supplemented by astrometry from Gaia EDR3 \citep[][]{GaiaCollab2021EDR3}, spectro-photometric distances from the AstroNN value-added catalogue for APOGEE \citep{Leung2019astronn, Leung2023measurement}, and \texttt{STARHORSE} \citep{Santiago2016spectro, Queiroz2018starhorse} based distances for PIGS \citep{Arentsen2024orbital}, resulting in 6D phase-space information for both surveys.

As we are interested in the inner 5 kpc region of the Galaxy, we will work in the Galactocentric coordinates ($X_{\rm GC}$, $Y_{\rm GC}$, $Z_{\rm GC}$), with cylindrical distance $R=\sqrt{X_{\rm GC}^2+Y_{\rm GC}^2}$. In this coordinate system, the Sun is placed at    $(X_{\rm GC},\,Y_{\rm GC},\,Z_{\rm GC})=(-R_{\odot},\,0,\,Z_{\odot})$, where $R_{\odot}=8.2$ kpc \citep{BlandHawthorn2016, Gravity2021, Leung2023measurement} and $Z_{\odot}=20.8\,{\rm pc}$  \citep{Bennett2019verical}. Besides Galactocentric Cartesian velocities $V_{X_{\rm GC}}$ and $V_{Y_{\rm GC}}$, we also use Galactocentric cylindrical radial $V_R$ and azimuthal $V_{\phi}$ velocities, where $V_{\phi}$ is defined positive in the direction of the Galactic rotation. For the transformation from the heliocentric to the Galactocentric Cartesian system, the position of the Sun $\left(R_{\odot},\, Z_{\odot}\right)$ and its Cartesian velocities $\mathbf{v_{\odot}}=\left(U_{\odot},\,V_0+V_{\odot},\,W_{\odot}\right)=(11.1\,{\rm km\,s^{-1}},\, 248.54\,{\rm km\,s^{-1}},\, 7.25\,{\rm km\,s^{-1}})$ are used, where $V_0$ is the circular velocity the Galaxy at the solar radius. The values for $U_{\odot}$ and $W_{\odot}$ are adopted from \citet*{Schoenrich2010local}, and for $V_0+V_{\odot}$ from \cite{Reid2020sagittarius} based on the proper motion of Sgr ${\rm A_{\ast}}$.  

\subsection{APOGEE: data release DR17}
\label{sec: dataApogee}
The APOGEE survey \citep{Majewski2016apogee}, a program within the Sloan Digital Sky Survey (SDSS), was designed to acquire high-resolution ($\sim 22500$) near-infrared ($1.5-1.7\,\mu$m) spectra of stars located in all major components of the Galaxy -- allowing APOGEE to make observations even in highly dust-extincted regions suitable for studying the inner Milky Way. 

The main survey target stars were randomly selected for observations from a high quality subset of the Two Micron All Sky Survey \citep[2MASS; ][]{Skrutskie2006two}, based on their $(J-K_s)_0$ colour and $H$-band apparent magnitude. For more information on the targeting strategies of APOGEE, see \cite{Zasowski2013ApogeeTarget, Zasowski2017ApogeeTarget}, \cite{Beaton2021final}, and \cite{Santana2021final}. Stellar parameters and chemical abundances for APOGEE stars are derived with the APOGEE Stellar Parameters and Chemical Abundances Pipeline \citep[ASPCAP;][]{Perez2016Aspcap, Holtzman2014Apogee, Johnson2022bdbs}. 
\begin{figure}
    \centering
    \includegraphics[scale=1.25]{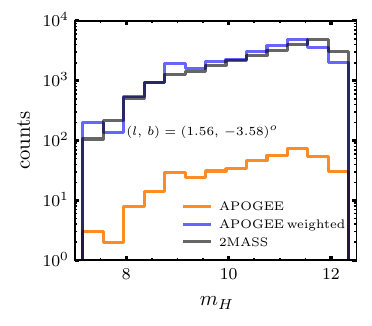}
    \caption{APOGEE luminosity function (apparent magnitude $m_H$) for a typical field close to the Galactic center before (orange) and after (blue) the photometric selection correction, respectively. The grey line shows the 2MASS luminosity function.}
    \label{fig: fig1}
\end{figure}

\begin{figure*} 
    \centering
    \includegraphics[scale=1.2]{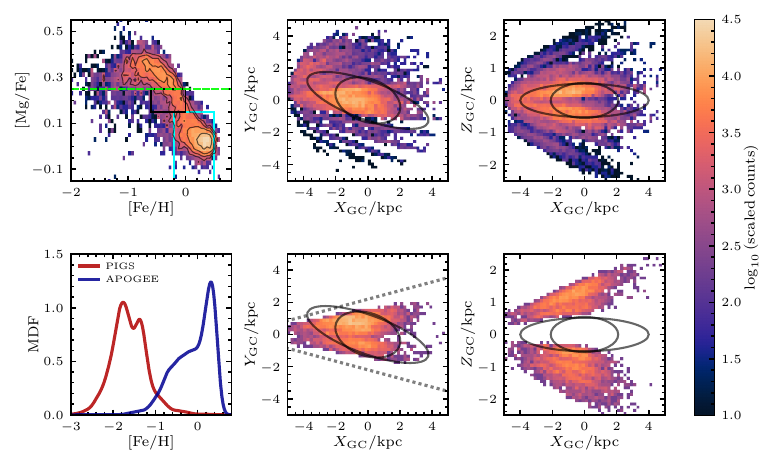}
    \caption{{\bf Top row:} {\bf First column:} The [Fe/H]-[Mg/Fe] distribution of APOGEE stars in our base sample (see section \ref{sec: dataApogee}). The green dashed line, the black and cyan rectangles show our definition of high-[Mg/Fe] ([Mg/Fe]$\geq0.25$), intermediate-[Mg/Fe] and low-[Mg/Fe] subsamples respectively. {\bf Second and third column:} The face-on and edge-on spatial distributions. The bigger and smaller grey ellipses are schematic representations of the Milky Way bar and bulge with the bar angle at $20^{\circ}$. 
    {\bf Bottom row:} {\bf First column:} Metallicity distribution of the PIGS and APOGEE stars. {\bf Second and third column:} Similar to the top row but for the PIGS stars. The two dotted lines show lines-of-sight at $\pm15^{\circ}$. APOGEE stars are weighted by their photometric selection weights, described in section~\ref{sec: dataApogee}. To ensure a common colour scale across all panels, each distribution is scaled so that the total count across all pixels is the same.}
    \label{fig: fig2}
\end{figure*}
From APOGEE DR17 \citep{Abdurro2022apogeeDR17}, we use [Fe/H] as metallicity, [Mg/Fe] as $\alpha$ abundance, line-of-sight velocities, and effective temperatures ($T_{\rm eff}$, used only for sample selection). These were complemented with proper motions from the Gaia EDR3 catalogue \citep{GaiaCollab2021EDR3}. Spectro-photometric distances are taken from the AstroNN value-added catalogue \citep{Leung2019astronn, Leung2023measurement} which are derived from a deep neural network trained on stars common to both APOGEE and Gaia. With the quality cuts mentioned below, the 50th and 84th percentiles of the AstroNN distance uncertainty distribution are approximately at $7\%$ and $12\%$ for stars in the bar region ($R\leq5$ kpc). 
To ensure reliable chemical abundances and precise distance estimates for stars in our sample, we impose the following additional criteria on the parent APOGEE DR17 catalogue:
\begin{itemize}
    \item Signal-to-noise ratio > 50
    \item \texttt{EXTRATARG} == 0
    \item no \texttt{${\rm Star\_Bad}$} flag: 23rd bit of \texttt{ASPCAPFLAG}==0
    \item $T_{\rm eff}>3200$ K
    \item astroNN distance errors $<20\%$
\end{itemize}

Finally, only stars with absolute galactic latitude $|b|<15^{\circ}$ and a cylindrical apocentre distance $R_{\rm apocentre}\leq5$ kpc\footnote{$R_{\rm apocentre}$ is calculated using the orbits of stars integrated in a dynamical Made-to-Measure potential model of the Milky Way, which includes a rotating bar with a pattern speed of $\Omega_{\rm b}=37.5\,{\rm km\,s^{-1}\,kpc^{-1}}$ \citep{Portail_2017}. For further details, refer to Section \ref{sec: potentialOrbits}.} are included. The latter criterion ensures that the stars are confined to the bar region, removing contamination from interlopers/halo stars. This leaves $22, 622$ APOGEE stars in our base sample.

In the bulge, the density falls off nearly exponentially along the bar axes \citep{Wegg2013mapping}. Moreover, using the 16th data release of APOGEE, \cite{Wylie2021a2a} reported a strong vertical metallicity gradient ($\sim-0.4\,{\rm dex\,kpc^{-1}}$) in the bar region \citep[see also][]{RojasArriagada2020many}. Since the APOGEE survey sample has nearly the same number of stars in fields at different latitudes, stars at greater heights are oversampled compared to those closer to the Galactic plane. Given the negative vertical metallicity gradient, this oversampling results in a higher proportion of metal-poor stars. To correct for this APOGEE selection effect, we follow a procedure similar to \cite{Bovy2014Apogee}. First, we obtain the colour and magnitude limits of each set of APOGEE stars observed together, referred to as {\it cohort}, and apply these limits to the 2MASS sample. Next, we calculate the ratio of the number of 2MASS stars to the APOGEE stars and assign it as a correction weight for each star in the cohort. Figure \ref{fig: fig1} illustrates the weighting scheme for a typical APOGEE field in the bulge region. These weights correct for the photometric selection fractions of the APOGEE survey and thereby provide first-order corrections for density and subsequent metallicity biases. They are used in constructing all subsequent figures except Figures~\ref{fig: fig6}, \ref{fig: fig7}, \ref{fig: figA2}.  

The top row of Figure \ref{fig: fig2} presents our base APOGEE sample. The first column shows the distribution of APOGEE stars in the [Fe/H]-[Mg/Fe] plane, confirming the chemical bimodality reported in previous studies of the Galactic bar/bulge region \citep{Rojas2019bimodal, Queiroz2020bule2disc, Wylie2021a2a}. The two dominant peaks arise at $\left({\rm [Mg/Fe],\,[Fe/H]}\right)\sim\left(0,\,0.3\right)$ and $\sim\left(0.35,\,-0.5\right)$. Several theoretical models for the chemical evolution in the central region are able to reproduce such a bimodality -- examples include the two-infall models \citep[][]{Chiappini1997_twoinfall, Spitoni2019, Spitoni2021}, early gas-rich merger models \citep[][]{Brook2004, Brook2007, Grand2018_bimodality}, and clump formation models \citep[][]{Immeli2004, Debattista2023}.

Due to the significant overlap of the two peaks, we select stars with ${\rm [Mg/Fe]}\geq0.25$ to obtain a cleaner sub-sample of high-[Mg/Fe] stars. Further, for a subsample of low-[Mg/Fe] stars in the Galactic bar and b/p-bulge, we select stars to have ${\rm[Mg/Fe]}\leq0.15$ and $-0.2\leq{\rm [Fe/H]}\leq0.5$. The upper limit of [Fe/H] is adopted to avoid contamination from the recently discovered lower rotating extremely metal-rich central knot of the Milky Way \citep{Rix2024EMRknot, Khoperskov2024III}. We also define a sub-sample of intermediate-[Mg/Fe] stars, by $-0.6<{\rm [Fe/H]}<0$ and $0.15<{\rm [Mg/Fe]}<0.25$ (black rectangle), to investigate the dependence of kinematics on [Mg/Fe] in the overlapping [Fe/H] range of low- and high-[Mg/Fe] stars.

The second and third columns in Figure~\ref{fig: fig2} display the spatial distributions of APOGEE stars in the face-on and the edge-on views, respectively. Notably, the APOGEE sample exhibits a near-far asymmetric coverage of the Galactic bar and bulge.

\subsection{PIGS: Pristine Inner Galaxy Survey}
\begin{figure*}
    \centering
    \includegraphics[scale=1.55]{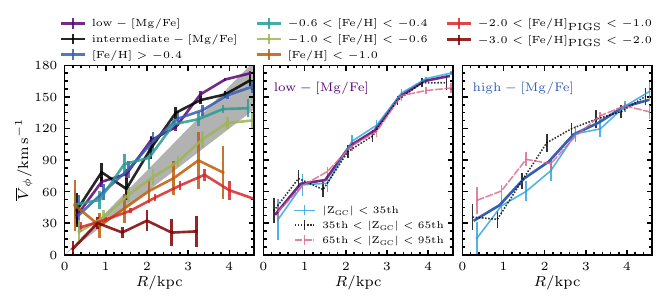}
    \caption{{\bf First column}: The mean Galactocentric rotational velocity $\overline{V}_{\phi}$ as a function of cylindrical radius $R$ for APOGEE and PIGS stars. The error bars are computed using the bootstrapping method, and $R$-bins with fewer than 10 stars are excluded. The low-[Mg/Fe] curve shows true $R$-bin positions, while other curves are alternately shifted left or right for clarity. The grey shade indicates the Milky Way bar's pattern speed range of $30-40\,{\rm km\,s^{-1}\,kpc^{-1}}$. {\bf Second and third column}: Cylindrical rotation pattern of $\overline{V}_{\phi}(R)$ for low- and high-[Mg/Fe] (${\rm [Fe/H]}>-1$) APOGEE stars. The $P$th percentiles of the $|Z_{\rm GC}|$ distribution over the entire radial range for low (high)-[Mg/Fe] sample are: ${\rm [35th,\,65th,\,95th]/kpc=[0.24,\,0.4,\,0.74]\left([0.32,\,0.54,\,1.2]\right)}$. The APOGEE stars are weighted by their photometric selection weights, also for computing the percentiles.}
    \label{fig: fig3}
\end{figure*}
\begin{figure*}
    \centering
    \includegraphics[scale=1.3]{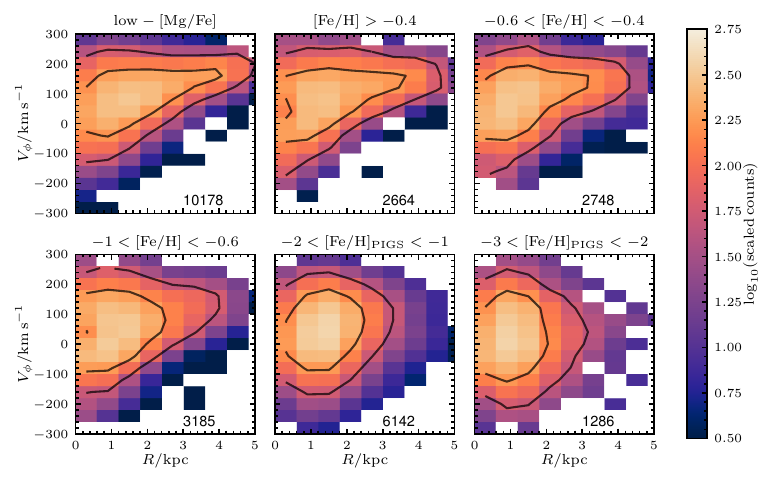}
    \caption{Distribution of APOGEE stars and PIGS stars in the $V_{\phi}-R$ plane. Each panel is scaled to maintain a constant total count across all pixels. The two contour lines in each panel are separated by 0.5 dex. The actual star count for each plot is given in its lower right corner. For the distributions, APOGEE stars are weighted by their photometric selection weights.}
    \label{fig: fig4}
\end{figure*}
PIGS is an extension of the main Pristine survey \citep{Starkenburg2017pristine}, targeting the metal-poor (${\rm [Fe/H]}\leq-1$) stars in the inner Galaxy (absolute longitude $|l|\leq12^{\circ}$ and absolute latitude $5^{\circ}<|b|<12^{\circ}$). Metal-poor candidates were pre-selected using metallicity-sensitive narrow-band CaHK photometry, see \cite{Arentsen2020pristine} for the target selection for PIGS. Of the metal-poor candidates in PIGS, 13000 were followed up from 2017 to 2020 with the Anglo Australian Telescope to obtain simultaneous low-resolution ($\sim 1300$) optical spectra ($3700-5500 \mathring{\rm A}$) and medium-resolution ($\sim 11000$) calcium triplet spectra ($8400-8800 \mathring{\rm A}$). These spectra were analysed with two independent full-spectrum fitting pipelines, \texttt{FERRE} and \texttt{ULySS}, to derive stellar parameters. In this work we use [Fe/H] from the FERRE stellar parameters, as recommended by \cite{Arentsen2020pristine}.

\cite{Arentsen2024orbital} derived distances for the PIGS stars using \texttt{STARHORSE} \citep{Santiago2016spectro, Queiroz2018starhorse}, a Bayesian isochrone matching method that estimates distances, extinction, and ages based on a combination of observables and priors. For the PIGS stars in the bar region ($R \leq 5$ kpc), the 50th and 84th percentiles of the resulting distance uncertainty distribution are approximately at $14\%$ and $20\%$, respectively.

\cite{Arentsen2024orbital} identified a subset of PIGS stars classified by \texttt{STARHORSE} as horizontal branch stars. These authors found that the horizontal branch stars appear to be more distant than red giant branch (RGB) stars and, as a result, excluded them from their analysis. However, we have verified that the $\overline{V}_{\phi}(R)$ profile for the horizontal branch stars is nearly the same as that for the RGB stars, so we keep them in our analysis for better statistics. Of the stars released by \cite{Arentsen2024orbital}, we select stars with $R_{\rm apocentre}\leq5$ kpc for our base PIGS sample, resulting in approximately 8000 stars.

The bottom row of Figure \ref{fig: fig2} shows the PIGS stars in our sample. The first column shows that the metallicity distributions of PIGS and APOGEE stars are complementary. The secondary peak in the metallicity distribution of the PIGS stars corresponds to the horizontal branch stars. The next two columns show the distributions of PIGS stars in the face-on and edge-on views -- completely missing the Galactic bar and bulge close to the Galactic plane ($|Z_{\rm GC}|\lesssim0.5$ kpc).

\section{Results}\label{sec: Results}

\subsection{\texorpdfstring{$\overline{V}_{\phi}-R$}{} profile: signature of the rotating bar in high-[Mg/Fe] stars}
\label{sec: kinematics}
To explore the origin of high-[Mg/Fe] stars in the bar region, we investigate their kinematics and compare them to low-[Mg/Fe] stars, believed to form the Galactic bar through the secular evolution of an early low-[Mg/Fe] thin disc \citep{Shen2010, Fragkoudi2018disc, Sanders2024bar}. The mean Galactocentric rotational velocity $\overline{V}_{\phi}$ for stars at any cylindrical distance $R$ that form a rotating bar with pattern speed $\Omega_{\rm b}$ is $\Omega_{\rm b}\times R$ plus the contribution from internal streaming motions \citep{Sellwood_1987, Vasquez_2013, Qin_2015, Tahmasebzadeh_2022}.

The second and third columns in Figure~\ref{fig: fig3} show that $\overline{V}_{\phi}\,(R)$ is nearly independent of the vertical height $|Z_{\rm GC}|$ for both low and high-[Mg/Fe] ([Fe/H] $<-1$) APOGEE samples, i.e., they follow a cylindrical rotation pattern. The cylindrical rotation pattern of bulge stars is already known in the heliocentric view, i.e., the line-of-sight velocity as a function of longitude does not change with latitude down to metallicity [Fe/H] $\sim-1$ \citep{Howard2009, Zocccali2014gibs, Ness2016ApogeeKinematics, Liao2024insights}. Therefore, in the first column in Figure~\ref{fig: fig3} we show $\overline{V}_{\phi}$ for the entire $|Z_{\rm GC}|$ distribution in radial bins, comparing high-[Mg/Fe] APOGEE and PIGS stars in different metallicity ranges, intermediate- and low-[Mg/Fe] APOGEE stars.

 Noteworthy points from the $\overline{V}_{\phi}-R$ curves shown in Figure \ref{fig: fig3} are that (i) in the bulge, high-[Mg/Fe] APOGEE stars with metallicity ${\rm [Fe/H]}>-0.6$ have, within errors, similar $\overline{V}_{\phi}$ as the low-[Mg/Fe] stars, and only 10-20\% lower values in the outer bar; (ii) $\overline{V}_{\phi}$ for intermediate-[Mg/Fe] stars agree in the bulge with values for both low- and high-[Mg/Fe] $\left({\rm [Fe/H]}>-0.6\right)$ stars, within errors, and fall between these in the outer bar; (iii) high-[Mg/Fe] stars with metallicity $-1.0<{\rm [Fe/H]}<-0.6$ show a linear increase in $\overline{V}_{\phi}$ with $R$, having slope consistent with their metal-rich counterparts but with reduced amplitude; and (iv) below metallicity ${\rm [Fe/H]}\sim-1$, both slope and amplitude decrease, consistent between APOGEE (orange line) and PIGS (red and maroon lines). 

The similarity in the rotational structure between low-[Mg/Fe] and the majority of high-[Mg/Fe] stars as well as its continuous transition with metallicity in high-[Mg/Fe] stars is further illustrated by the distribution of the different stellar samples in the $V_{\phi}-R$ plane in Figure \ref{fig: fig4}. The distribution of high-[Mg/Fe] metal-rich (${\rm [Fe/H]}>-0.4$) stars is very similar to that of low-[Mg/Fe] stars. With decreasing metallicity, the slopes of the lower boundaries of the contours do not notably change but the velocity dispersion at a given radial bin increases progressively and the radial distribution shrinks. \cite{Wylie2021a2a} reported similar results with heliocentric radial velocities. For the PIGS stars, covering the metallicity range ${\rm [Fe/H]}<-1$, the contours are roundish and radially concentrated, showing primarily dispersion-dominated kinematics \citep{Arentsen2020pristine, Rix2022poor}.

These results show that the $V_{\phi}-R$ distribution of the metal-rich (${\rm [Fe/H]}\geq-0.6$) high-[Mg/Fe] stars is similar to that for the bar-driven low-[Mg/Fe] stars, with only mildly increased velocity dispersion, while the transition to mostly dispersion-dominated kinematics occurs only at low metallicities. This indicates that high-[Mg/Fe] stars down to ${\rm [Fe/H]}\sim-0.6$ are part of the Galactic bar and have resulted from secular evolution of an early high-[Mg/Fe] disc.

\subsection{The mean radial velocity fields}
\begin{figure*}
    \centering
    \includegraphics[scale=1.35]{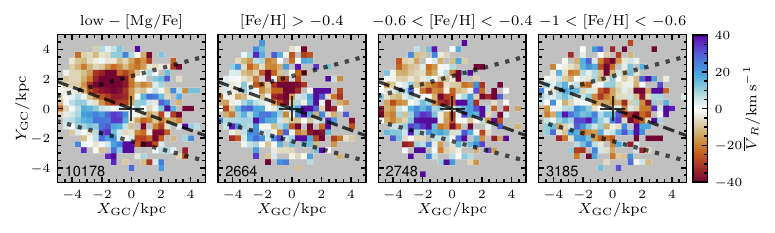}
    \caption{Observed mean radial velocity $\overline{V}_R$ fields for APOGEE samples as in Figure~\ref{fig: fig3}. The velocity fields are smoothed using a 2D Gaussian kernel with a width of 0.625 kpc in each dimension (pixel size is $0.4\times0.4\,{\rm kpc^2}$). The black dashed line is inclined at $20^{\circ}$, and the `+' symbol marks the Galactic center. Two dotted lines indicate lines of sight at $\pm15^{\circ}$. The actual star count for each plot is given in its lower left corner. APOGEE stars are weighted by their photometric selection weights.}
    \label{fig: fig5}
\end{figure*}
To complement the results from the rotational component of the velocity, here we investigate the mean Galactocentric radial velocity $\overline{V}_R$ fields of APOGEE stars in the bar region. The mean radial velocity field is known to exhibit a quadrupole pattern around the center, a signature of bars seen in the Milky Way observations and $N$-body simulations \citep{Bovy2019life, Queiroz2021milky, Leung2023measurement, GaiaDr3_2023, Zhang2024kinematics} as well as in cosmological simulations \citep{Fragkoudi2020chemodynamics}. The quadrupole pattern arises due to the streaming motion of bar-supporting stars along the elongated orbits. 

Figure~\ref{fig: fig5} shows the mean radial velocity fields for the APOGEE stellar samples as in Figure~\ref{fig: fig3}. In the case of the low-[Mg/Fe] stars (leftmost column), the two poles on the nearer side of the bar can be clearly noticed; however, the corresponding poles on the farther side are weaker owing to the poorer sampling, see Figure \ref{fig: fig2}. The subsequent panels show high-[Mg/Fe] stars in intervals of decreasing metallicity as indicated. Even with the lower statistics ($\sim2500$ stars) compared to the low-[Mg/Fe] sample, the two poles on the nearer side of the bar can be noticed in the metallicity ranges ${\rm [Fe/H]}>-0.4$ and $-0.6<{\rm [Fe/H]}<-0.4$. However, in the rightmost column ($-1<\rm{[Fe/H]}<-0.6$) the two poles are extended beyond the Galactic center, thereby making the quadrupole pattern unclear. 

Coherent with the rotational properties, the metal-rich (${\rm [Fe/H]}\geq-0.6$) high-[Mg/Fe] stars exhibit a quadrupole pattern in the mean radial velocity field similar to the low-[Mg/Fe] stars, which becomes weaker and more centrally concentrated at lower metallicity.

\subsection{Frequency analysis of orbits}
\label{sec: orbitalFreq}
The previous two sections show that the observed kinematic properties of the metal-rich (${\rm [Fe/H]}\geq-0.6$) high-[Mg/Fe] stars in the bar region are similar to those of the co-spatial, bar-participating low-[Mg/Fe] stars, with a gradual transition to non-bar-like properties at lower metallicities. These findings not only confirm that the majority of high-[Mg/Fe] stars are part of the Galactic bar but also suggest that they originated from the secular evolution of an early high-[Mg/Fe] disc. Now we aim to investigate the metallicity dependence of bar-like properties using the orbital information of individual stars. The goal is to categorize them into the key orbit families that arise in a rotating barred potential \citep{Contopoulos1980, Voglis_2007, Valluri2016unified}, aiming to explore the relative contribution of these orbit families to the overall rotation as a function of metallicity.

\subsubsection{Potential models and orbit integration}
\label{sec: potentialOrbits}

\begin{figure*}                                                       
    \centering
        \includegraphics[scale=1.6]{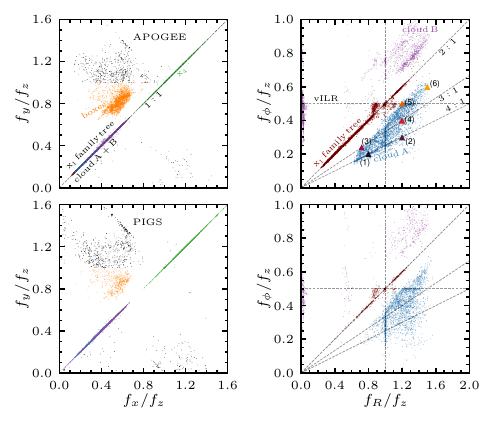}
    \caption{{\bf Top row:} Distribution of APOGEE stars in Cartesian (left) and cylindrical (right) frequency ratio planes for our standard potential model (see section \ref{sec: potentialOrbits}). Maroon, orange, blue, purple, green, and black scattered points represent the ${\rm x_1\,family\,tree,\,boxes,\,cloud\,A,\,cloud\,B,\,x_4,\,and\,``others"}$ orbit families. The ILR (Inner Lindblad Resonance, ${\rm f_R:f_{\phi}}=2:1$), $3:1$, $4:1$, and ${\rm vILR}$ (vertical ILR, $f_{\phi}/f_z=0.5$) resonances are indicated using grey-dashed lines in the right panel. The orbits lying on the left of the line $f_R/f_z=1$ extend radially to the outer bar and are thinner, while those on the right are more in the bulge and thicker. Triangular points of varying colours labelled (1) to (6) mark typical orbits from the ${\rm cloud\,A}$ family, which are shown in Figure~\ref{fig: figA2}. {\bf Bottom row:} Same as top row, but for PIGS stars.}
    \label{fig: fig6}
\end{figure*}

For each APOGEE and PIGS star in our sample, we treat the observed phase space coordinates as initial conditions and integrate its orbit in a potential from the suite of dynamical bar-bulge models developed by \cite{Portail_2017}. These models were adapted using the Made-to-Measure method to fit bulge density and kinematic data from the VVV, UKIDSS, 2MASS, BRAVA, OGLE, and ARGOS surveys. In these models, based on the distribution of red clump stars from \cite{Wegg2013mapping} and \cite{Wegg_2015}, the distance from the Sun to the Galactic center is  $R_{\odot}=8.2$ kpc with a bar angle $\alpha_{\odot}=28^{\circ}$.

As the standard case, we use a potential model with mass to red clump star ratio of 1000 $M_{\odot}$, central in-plane mass $M_c = 2 \times 10^9\,M_{\odot}$ , and pattern speed of $\Omega_{\rm b} = 37.5\,{\rm km\,s^{-1}\,kpc^{-1}}$ \citep[see][]{Portail_2017}. This model provides a good fit to both the bulge proper motions \citep{Clarke2019virac} and the inner Galaxy gas dynamics \citep{Li2022gas}. Moreover, an analytic approximation of this model is available from \cite{Sormani_2022} in the \texttt{AGAMA} library \citep{Vasiliev_2019}.

To ensure that our conclusions do not depend on a specific potential model, we also use an alternative model from \cite{Portail_2017} differing in the shape of potential. In this alternative model, the mass to red clump star number is 1100, the central in-plane mass is half compared to the standard model, and the bar pattern speed is $\Omega_{\rm b} = 35\,{\rm km\,s^{-1}\,kpc^{-1}}$. 

In Figure \ref{fig: fig5} we found that the quadrupole moment in the mean radial velocity fields of observed stars is inclined at $\sim20$ degrees \citep[][]{Leung2023measurement, GaiaDr3_2023}. This could suggest (i) a slightly lower bar angle $\alpha_{\odot}$, (ii) bending of the bar ends by the torques from the spiral arms \citep[e.g.][]{Valpuesta_2011}, (iii) presence of spiral arms stars in the outer bar region \citep{Hilmi_2020}, or (iv) apparent stretching of the mean radial velocity field by distance errors \citep{Gomez_2015, Vislosky2024gaiadr3}. 

Using an approach similar to \cite{Zhang2024kinematics}, we found that if the AstroNN distance errors are as small as given, their effect is not sufficient to explain the difference between the model value of $\alpha_{\odot}=28^{\circ}$ and a value around $20^{\circ}$. Therefore we used a third potential model with the Galactic bar inclined at $22^{\circ}$. 

We computed orbits in all three potentials and confirmed that our conclusions remain unchanged, see section \ref{sec: sec3_4} and Figure~\ref{fig: figA1}. For each case, we transformed the sample stars from the heliocentric to the rotating bar frame, and integrated orbits using the drift-kick-drift leap-frog integration algorithm of the NMAGIC code \citep{DeLorenzi2007nmagic} for 2 Gyr and saving orbit points at every 1 Myr. 

\subsubsection{Frequency maps}
For each orbit, we define the cylindrical apocentre distance $R_{\rm apocentre}$ as the $90$th percentile of the time series for the cylindrical distance $R$. For all orbits, we compute Cartesian coordinate orbital frequencies ($f_x$, $f_y$, $f_z$) as well as cylindrical radial and azimuthal frequencies $(f_R,\,f_{\phi})$ in the bar frame. To compute an orbital frequency, say $f_x$, we compute the fast Fourier transform of its time series along the orbit and identify $f_x$ as the frequency corresponding to the highest spectral peak. 

Figure \ref{fig: fig6} shows the distribution of APOGEE and PIGS stars with cylindrical apocentre distance $R_{\rm apocentre}\leq5$ kpc in Cartesian and cylindrical frequency ratio planes. Following \cite{Voglis_2007} and \cite{Valluri2016unified}, we use these frequency ratios to classify orbits into the main orbit families that arise in a rotating barred potential -- ${\rm x_1}$ family tree; boxes; ${\rm x_4}$; ${\rm cloud\,A}$ and ${\rm B}$. To separate the retrograde ${\rm x_4}$ orbits from prograde orbits, we also use the mean rotational velocity $\hat{V}_{\phi}$ along the orbits. The loci of these orbit families are indicated in Figure~\ref{fig: fig6} and the corresponding criteria in Table \ref{tab: table1}. 

Note that the ${\rm x_1}$ family tree, ${\rm cloud\,A}$ and ${\rm B}$ families occupy similar loci (1:1 line) in the Cartesian frequency ratio plane; therefore, we use the cylindrical frequency ratio to break this degeneracy. The ${\rm cloud\,A}$ includes higher order resonant orbit families (3:1, 4:1), some that support the bar and are boxy in shape, and others with spherical orbits -- Figure \ref{fig: figA2} shows typical orbits from the ${\rm cloud\,A}$ family across its frequency range, marked with triangular points and numbers in Figure \ref{fig: fig6}. The ${\rm cloud\,B}$ orbits are sub-dominant across the metallicity range of APOGEE and PIGS (see Figure \ref{fig: fig7}), and do not contribute to the bar and boxy/peanut bulge. 

Figure~\ref{fig: figA3} shows the distribution of different orbit families for both APOGEE and PIGS stars in the $R_{\rm apocentre}-Z_{\rm max}$ plane: APOGEE orbits associated with the ${\rm x_1}$ family tree and ${\rm cloud\,A}$ orbits have flattened shapes and extend radially to the outer bar, whereras boxes and ${\rm x_4}$ orbits are mostly confined to the bulge and are more spheroidal. The PIGS star ${\rm x_4}$ orbits have nearly spherical shapes.

While the distributions of APOGEE and PIGS orbits that belong to the ${\rm x_1}$ family tree and ${\rm cloud\,A}$ appear similar in the Cartesian frequency ratio plane, they differ in the cylindrical frequency ratio plane -- PIGS orbits occupy mainly the $\frac{f_R}{f_z}\geq1$ part of the plane and are localized in the inner 3 kpc region with considerable vertical extension. 

The ${\rm x_4}$ orbit family lies on the 1:1 line in the Cartesian frequency ratio plane with a higher frequency ratios $f_x/f_z$ and $f_y/f_z$. These are short-axis tubes with a retrograde sense of rotation in the rotating bar frame \citep[][]{Contopoulos1980, Valluri2016unified}. Additionally, we have found a very small fraction of ${\rm x_2}$ orbits -- less than $0.05\%$, in both APOGEE and PIGS samples. This minimal presence of the ${\rm x_2}$ orbit family is consistent with the orbital analysis in $N$-body bar models \citep{Sellwood_1987, Voglis_2007}. Consequently, we do not include this orbit family in our analysis. 

Finally, some orbits do not belong to any of the major families, shown as black scattered points in the left columns of both rows in Figure~\ref{fig: fig6}. We refer to these orbits as "others". These are more frequent but still subdominant in the PIGS orbit sample.

\begin{table}
  \begin{center}
    \caption{Orbit classification using frequency ratios and $\hat{V}_{\phi}$. For each orbit, $\hat{V}_{\phi}$ denotes the mean of the bar frame rotational velocity along the orbit. The `--' symbol indicates that no selection criterion is applied for the corresponding quantity.}
    \label{tab: table1}
    \begin{tabular}{ |c|c|c|c|c|c| }
        \hline
        Family & $\frac{f_x}{f_z}$ & $\frac{f_y}{f_z}$ & $\frac{f_y}{f_x}$ & $\frac{f_R}{f_{\phi}}$ & $\hat{V}_{\phi}$\\ 
        \hline
        \hline
        ${\rm x_1}$ family tree & $\leq0.7$ & -- & $=1\pm0.1$ & $=2\pm0.1$ & --\\
        \hline
        boxes & -- & $\in(0.6, 1)$ & $>1.1$ & -- & --\\
        \hline
        ${\rm cloud\,A}$ & $\leq0.7$ & -- & $=1\pm0.1$ & $>2.1$ & --\\
        \hline
        ${\rm cloud\,B}$ & $\leq0.7$ & -- & $=1\pm0.1$ & $<1.9$ & --\\
        \hline
        ${\rm x_4}$ & $>0.7$ & -- & $=1\pm0.1$ & -- & $<0$\\
        \hline
        ${\rm x_2}$ & $>0.7$ & -- & $=1\pm0.1$ & -- & $>0$\\
        \hline
    \end{tabular}
  \end{center}
\end{table}

\subsection{Fraction of orbit families as a function of metallicity}
\begin{figure}
    \centering
    \includegraphics[scale=1.225]{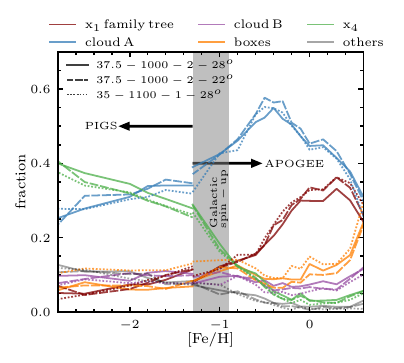}
    \caption{Metallicity dependence of the fraction of stars associated with different orbit families. Line styles specify potential models: pattern speed (${\rm km\,s^{-1}\,kpc^{-1}}$)--mass to red clump stars ratio--central in-plane mass ($10^{10}\,M_{\odot}$)--bar angle (degrees); see section \ref{sec: potentialOrbits} and \protect\cite{Portail_2017}. The grey stripe marks the metallicity range of the Galactic spin-up identified in the solar-neighbourhood analysis \protect\citep{Belokurov2022Dawn}.}
    
    \label{fig: fig7}
\end{figure}

{\label{sec: sec3_4}}
After classifying APOGEE and PIGS stars into the key orbit families, Figure~\ref{fig: fig7} shows the fraction of each orbit family as a function of metallicity. In addition to our standard potential model, the orbit fractions for two other potential models as described in section \ref{sec: potentialOrbits} are over-plotted. As metallicity decreases from ${\rm [Fe/H]}\sim+0.3$ to ${\rm [Fe/H]}\sim-0.4$, the fraction of ${\rm x_1}$ family tree orbits declines steadily from approximately $35\%$ to $20\%$. These metallicity values correspond to the peaks of the metallicity distributions for the low- and high-[Mg/Fe] sequences. The fraction of ${\rm x_1}$ family tree orbits falls below $10\%$ at metallicity $\lesssim -1$. 

Additionally, the fraction of boxes decreases sharply from approximately $25\%$ at metallicity ${\rm [Fe/H]}\sim0.6$ to about $10\%$ at ${\rm[Fe/H]}\sim0.2$, and remains nearly constant at lower metallicities. Box orbits are particularly prominent in the bulge, around 1 kpc from the center (Figure~\ref{fig: figA3} in the appendix). Recently, \cite{Rix2024EMRknot}  examined the spatial distribution of stars with GAIA XP spectra, and reported an overall metal-rich (predominantly ${\rm[M/H]_{XP}}\gtrsim0.5$) compact spheroidal knot at $R\leq1.5$ kpc \citep[see also][]{Khoperskov2024III}.

Figure \ref{fig: fig7} shows another interesting result. In the low metallicity range, [Fe/H] = (-3, $\sim-1.3$), stars associated with ${\rm x_4}$ and ${\rm cloud\,A}$ orbits dominate, with their fractions varying by $\lesssim 15\%$ over this range. However, as metallicity increases further from [Fe/H] $\sim -1.3$ to $\sim -0.9$, the fraction of stars associated with ${\rm cloud\,A}$ orbits rises sharply, while the fraction of counter-rotating ${\rm x_4}$ orbits shows a pronounced decline. In this metallicity range, also the fraction of stars associated with the ${\rm x_1}$ family tree begins to increase, reaching a plateau around solar metallicity. As shown in the next section, this is a signature of the Galactic spin-up, already seen in the solar neighbourhood \citep{Belokurov2022Dawn} and inwards \citep{Rix2022poor}.

\subsection{Rotational velocity as a function of metallicity: "spin-up" in the bar region}
\begin{figure*}
    \centering
    \includegraphics[scale=1.3]{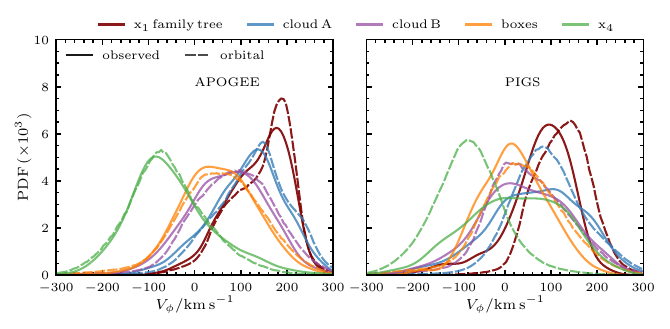}
    \caption{The Galactocentric rotational velocity distributions for the observed stars associated with different orbit families (solid lines), and their corresponding orbit-based velocity distributions (dashed lines; see text for details). The left and right panels show APOGEE and PIGS stars, respectively, with $R_{\rm apocentre}<5$ kpc. APOGEE stars are weighted by their photometric selection weights.}
    \label{fig: fig8}
\end{figure*}

Metal-rich (${\rm [Fe/H]}\gtrsim-0.6$) stars in the bulge are known to be rapidly rotating. On the other hand, recent studies of metal-poor (${\rm [Fe/H]}\lesssim-1.5$) stars in the inner regions have shown that their kinematics is dominated by dispersion with a small, positive rotation. These stars have been associated with the proto-Galactic population \citep[][]{Rix2022poor, Arentsen2024orbital}. The reason for the small prograde rotation at such low metallicity is not clear yet. Therefore, in addition to the overall rotational velocity, we also investigate the contribution of each orbit family as a function of metallicity.

\subsubsection{Rotational properties of orbit families}\label{sec: Sec_3_5_1}
 For stars belonging to a given orbit family in the APOGEE or PIGS surveys, Figure~\ref{fig: fig8} presents both their distribution of rotational velocities from the observed phase-space coordinates, and their combined distributions of rotational velocities sampled along their orbits, leading to a single `orbit-based' velocity distribution for the family. Both distributions are computed in the Galactocentric reference frame. Recall from section \ref{sec: potentialOrbits} that each orbit is integrated for 2 Gyr and its phase-space coordinates are saved at every 1 Myr. Rotational velocities at each saved orbit point for all orbits of the family enter the orbit-based distribution.
 
The observed and orbit-based rotational velocity distributions agree well for all the orbit families in the APOGEE sample and generally for the PIGS sample, except for the ${\rm x_4}$ orbit family. This agreement indicates that our potential model is a good approximation of the Galactic bar. 

For the ${\rm x_4}$ orbit family, the PIGS orbit-based rotational velocity distribution matches that in APOGEE and is retrograde as expected, with a  peak at $V_{\phi}\sim-90\,{\rm km\,s^{-1}}$. However, the observed distribution for the PIGS ${\rm x_4}$ stars peaks near $V_{\phi}\sim0\,{\rm km\,s^{-1}}$ and has a broader shape. We noticed there is an unusual overdensity of prograde velocity PIGS stars linked to the ${\rm x_4}$ orbit family near the $y-z$ plane of the bar (persisting also when excluding the horizontal branch stars from the PIGS sample). Examining their orbits shows that these stars are indeed more likely to attain prograde velocities when they are near the $y-z$ plane of the bar and close to their maximum vertical extent, but this occurs only rarely along their orbits. We investigated the origin of this overdensity in the Appendix, Figure~\ref{fig: figA6}.  This demonstrates that by applying the spatial selection of the PIGS survey in latitude and distance to the APOGEE stars in a metallicity range common to both surveys and convolving with a Gaussian distance error distribution corresponding to the larger distance errors in PIGS, the curious observed rotational velocity distribution of ${\rm x_4}$ stars can be well-reproduced.

The left panel of Figure \ref{fig: fig8} shows that the APOGEE ${\rm x_1}$ family tree and ${\rm cloud\,A}$ orbit families are associated with a prograde velocity distribution, exhibiting peaks at $\geq 150,{\rm km/s}$ and are skewed toward lower velocities. Further, Figure \ref{fig: figA4} shows that the resonant $\rm cloud\,A$ orbits that are confined within the Galactic bulge have lower mean rotation and larger velocity dispersion compared to those extending to the outer bar region. In contrast, the velocity distributions for the boxes and the ${\rm x_4}$ families are nearly symmetric, with peaks at approximately $50$ and $-90\,{\rm km\,s^{-1}}$, respectively, and a dispersion of approximately $ 80\,{\rm km\,s^{-1}}$ in both cases. In summary, the ${\rm x_1}$ family tree, ${\rm cloud\,A}$, and box families are associated with prograde rotation, where the mean rotation decreases and the velocity dispersion increases progressively from the ${\rm x_1}$ family tree to the box orbits. Conversely, the ${\rm x_4}$ family is associated with retrograde rotation.

\begin{figure}
    \centering
    \includegraphics[scale=1.225]{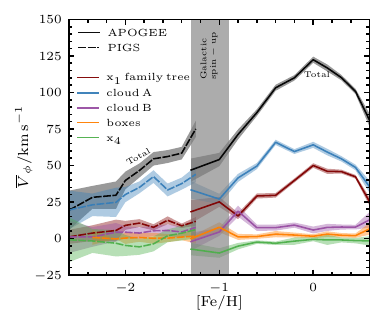}
    \caption{Metallicity dependence of the mean observed Galactocentric rotational velocity for the APOGEE (solid lines) and  PIGS stars (dashed lines). Further, contributions of stars from different orbit families to the overall rotational velocity (black lines) are shown using line colours as indicated. Stars with $1.5\leq R/{\rm kpc}\leq3.5$ are included. The error bars, shaded areas, are computed as $\sigma_{V_{\phi}}/\sqrt{N}$, where $\sigma_{V_{\phi}}$ and $N$ denote the rotational velocity dispersion and actual number of stars in each metallicity bin. The APOGEE stars are weighted by their photometric selection weights.}
    \label{fig: fig9}
\end{figure}

\begin{figure*}
    \centering
    \includegraphics[scale=1.55]{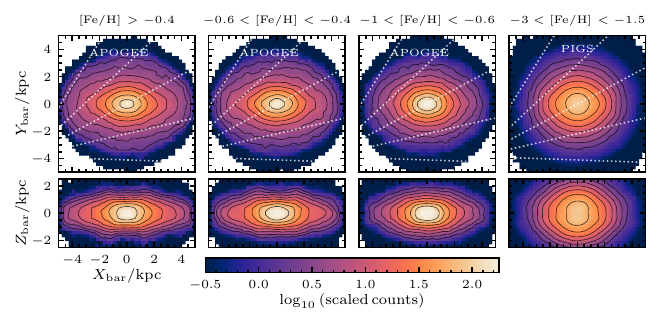}
    \caption{Orbital density distributions in the bar frame for high-[Mg/Fe] APOGEE and PIGS stars in our standard potential model. Top row: Face-on view, including orbital points with $|Z_{\rm bar}|<1$ kpc. Bottom row: Edge-on view, including points with $|Y_{\rm bar}|<1$ kpc. Each panel is scaled to maintain a constant total count across all pixels. Consecutive iso-density contours represent a density change of 0.25 dex. Dotted white lines indicate lines of sight from the Sun at  $0,\,\pm15^{\circ},\,\pm30^{\circ}$. APOGEE stars are weighted by their photometric selection weights.}
    \label{fig: fig10}
\end{figure*}

Figure~\ref{fig: figA5} additionally presents the angular velocity $\left(V_{\phi}/R\right)$ distributions of low- and high-[Mg/Fe] APOGEE and PIGS stars grouped by their orbit families. The ${\rm x_1}$ family tree and cloud A are the dominant families in both APOGEE samples (see also Figure~\ref{fig: fig7}). The peaks of their angular velocity distributions align with the expected bar pattern speed, with a slightly shifted peak for the ${\rm x_1}$ family tree due to larger streaming motions. In contrast, PIGS star orbits are predominantly cloud A and ${\rm x_4}$ orbits. The former exhibit a significantly broader distribution despite a peak consistent with the bar pattern speed, while the latter orbit family has a retrograde peak with an even wider distribution. These results again illustrate the similarity in the orbital dynamics between the low- and high-[Mg/Fe] stars, with a notable difference from the more metal-poor PIGS sample (${\rm [Fe/H]}\lesssim-1$).

\subsubsection{Orbital contribution to the overall rotational velocity}
We have now quantified the fraction of stars belonging to each orbit family as a function of metallicity (Figure \ref{fig: fig7}) as well as their rotational properties (Figure \ref{fig: fig8}). Although Figure \ref{fig: fig8} shows the rotational velocity distributions without binning in metallicity, the distributions do not change significantly across the metallicity range for APOGEE and PIGS samples. Figure~\ref{fig: fig9} shows the mean observed rotational velocity (in the Galactocentric frame) as a function of metallicity, and coloured lines show the contribution of stars associated with each orbit family to the total rotational velocity. The mean rotation at any metallicity can be expressed as $\overline{V}_{\phi}=\sum_{i}\left(w_i*\overline{V}_{\phi,\,i}\right)$, where $w_i$ is the fraction of stars that belong to the $i$-th orbit family and $\overline{V}_{\phi,\,i}$ is their mean rotation velocity, for each metallicity.

Note that to construct Figure~\ref{fig: fig9}, only the label (${\rm x_1}$, etc.) for each star comes from the orbital information, while the $V_{\phi}$ and $R$ come from the observed 6D phase-space. Further, following \cite{Arentsen2024orbital}, only stars with $1.5\leq R/{\rm kpc}\leq3.5$ and $R_{\rm apocentre}\leq 5$ kpc are included. 

The overall rotational velocity for APOGEE stars (solid black line) shows a sharp rise as metallicity increases from ${\rm [Fe/H]}\sim-1.3$ to solar metallicity -- indicating a spin-up in the bar region similar to that observed at larger radii \citep{Belokurov2022Dawn, Rix2022poor}. We note that this result is mostly driven by the stars outside the bulge whereas the rise of rotation velocity in the bulge is more gradual. Stars associated with the ${\rm x_1}$ and ${\rm cloud\,A}$ orbit families contribute most to the overall rotational velocity for ${\rm [Fe/H]}\gtrsim-1$.

Additionally, below metallicity ${\rm [Fe/H]}\sim-0.6$, stars associated with the ${\rm x_4}$ orbit family show net retrograde rotational velocity, see also Figure \ref{fig: fig8} -- this means that these stars are even more retrograde in the rotating bar frame. This is interesting in itself as even under the influence of the prograde rotating Galactic bar, these stars have managed to maintain mean retrograde rotation. 

For PIGS stars covering the metallicity range of $-3.0\lesssim{\rm[Fe/H]}\lesssim-1.3$, the overall mean rotation decreases from $V_{\phi}\sim75\,{\rm km\,s^{-1}}$ at ${\rm [Fe/H]}\sim-1.3$ to $V_{\phi}\sim35\,{\rm km\,s^{-1}}$ at ${\rm [Fe/H]}\sim-2.2$, and then becomes flat at still lower metallicities \citep[see also][]{Arentsen2024orbital}. Across the entire metallicity range for PIGS, only stars linked to the ${\rm cloud\,A}$ orbit family contribute significantly to the overall rotational velocity in Figure~\ref{fig: fig9}, but see the further discussion in Section~\ref{sec: sec4_2}. Note that the ${\rm cloud\,A}$ orbits of the PIGS stars are relatively flattened orbits, as shown in Figure~\ref{fig: figA3}.

\subsection{Orbital density distributions}
The previous sections show the interplay of the orbit families as a function of metallicity, revealing a continuous transition in their fractions as a function metallicity. In this section, orbital density distributions and their metallicity dependence are presented.

Figure \ref{fig: fig10} shows the orbital density distributions for high-[Mg/Fe] APOGEE stars in each of three metallicity ranges down to ${\rm [Fe/H]}\sim-1$, and for PIGS stars with [Fe/H] $<-1.5$. For the high-[Mg/Fe] APOGEE stars with [Fe/H] $>-0.4$ (left-most column), the face-on view shows a clear bar (especially in the fourth and fifth contours) and the edge-on view reveals a strong peanut shaped structure in the central 2 kpc region. In the next metallicity range ($-0.6<{\rm [Fe/H]}<-0.4$), the face-on view still shows a bar similar to the higher metallicity range, but the peanut is thicker. For $-1.0<{\rm [Fe/H]}<-0.6$, the face-on distribution appears more oval-shaped, and the edge-on view becomes boxier and even thicker. In the right-most column, the orbital density distribution for PIGS stars is nearly spheroidal.

These orbital density distributions are constructed using orbits that are integrated in a realistic potential model starting from the observed initial conditions for each star.  Figure \ref{fig: figA1} shows that the orbital density distributions of low- and high-[Mg/Fe] stars are consistent across the different potential models described in section \ref{sec: potentialOrbits}. One caveat here is the spatial selection of surveys arising due to the sparse distribution of fields or missing fields, especially at high latitudes ($|b|>5^{\circ}$). However, the photometric selection function correction weights used for APOGEE stars would correct for most of the vertical bias.

Recently, \cite{Khoperskov2024III} used a novel orbit superposition approach combined with the APOGEE DR17 to approximately correct for the survey selection function. Briefly, they computed orbits for the APOGEE stars in the barred potential from \cite{Sormani_2022}, which is an analytic model based on the potential constructed by \cite{Portail_2017} used as our standard potential here, with the same pattern speed. They adjusted the orbital weights of observed stars to match the 3D stellar density of the model \citep[for details, see][]{Khoperskov2024II, Khoperskov2024I}. The orbital density distributions shown in Figure~\ref{fig: fig10} agree qualitatively with \cite{Khoperskov2024III} (their figure 17), suggesting that the correction for the photometric selection of APOGEE, as described in section~\ref{sec: dataApogee}, effectively accounts for the most important selection effects.

 In summary, the orbital density distributions show a prominent but thick bar with peanut bulge for high-[Mg/Fe] APOGEE stars with [Fe/H] $>-0.6$. As metallicity decreases further to -1, the face-on distribution becomes oval-shaped with a boxy bulge in the edge-on view. At even lower metallicities, the distribution transitions to a spheroidal shape. These results align with the model-independent dynamical analysis of the stellar samples presented in earlier sections.

\section{Discussion}
\label{sec: Discussion}

\subsection{Dynamics of high-[Mg/Fe] stars in the bar region}\label{sec: 41}
Figure~\ref{fig: fig3} showed a cylindrical rotation pattern in the mean Galactocentric rotational velocity $\overline{V}_{\phi}(R)$ for both low- and high-[Mg/Fe] APOGEE stars across the bulge and the outer bar. This is consistent with simulations following thin and thick discs which form a bar and boxy/peanut bulge: In their $N$-body model, \cite{Fragkoudi2017} found that both thin and thick disc stars separately, and also all stars together, exhibit cylindrical rotation in their line-of-sight velocities, and \citet{Debattista2017fractionation} found a similar result for stars of different ages in their star forming simulation.

Figure~\ref{fig: fig3} further showed that the mean rotational $\overline{V}_{\phi}(R)$ for high-[Mg/Fe] stars down to metallicity [Fe/H] $\sim-0.6$ is consistent within errors with that of low-[Mg/Fe] stars in the bulge, and is only $10-20\%$ lower in the outer bar. Also the $V_{\phi}-R$ distributions of metal-rich ([Fe/H] $>-0.6$) high-[Mg/Fe] stars are similar to those of low-[Mg/Fe] stars in the bulge (Figure~\ref{fig: fig4}), but with increased dispersion especially in the outer bar. 

A dispersion-dominated system, such as a classical bulge, can acquire angular momentum from the rotating bar and display cylindrical rotation \citep{Saha_2012, Saha_2013, Saha_2016}. However, cylindrical rotation was found to extend beyond the half-mass radius only in the low-mass classical bulge ($\sim 1/5$ of total final b/p bulge mass) from \citet{Saha_2012}, but was not seen in more massive spun-up classical bulges \citep{Saha_2016}. The angular momentum transfer generates a characteristic linear rotational velocity profile in the classical bulge, with $\overline{V}_{\phi}/\sigma_{V_{\phi}}\sim0.3$ at two half-mass radii of the classical bulge in the low-mass classical bulges and $\sim 0.2$ in the more massive ones, significantly lower than for the b/p bulges in these simulations. The rotation profiles in Figure~\ref{fig: fig3} correspond to $\overline{V}_{\phi}/\sigma_{V_{\phi}}\simeq 0.9$, and $0.6$ at $R=1.5$ kpc for the metal-intermediate $({\rm [Fe/H]}>-0.6)$ and metal-poor $(-1<{\rm [Fe/H]}<-0.6)$ high-[Mg/Fe] APOGEE stars, respectively\footnote{For comparison, $\overline{V}_{\phi}/\sigma_{V_{\phi}}$ at this radius reaches $0.9$ for the low-[Mg/Fe] stars, and $0.4$ and $0.2$ for PIGS stars in the metallicity ranges $-1.5<{\rm [Fe/H]}<-1$ and $-2<{\rm [Fe/H]}<-1.5$, respectively.}. This argues that most of the high-[Mg/Fe] bulge rotates too rapidly to be explained as a spun-up classical bulge.

In the mean radial velocity fields presented in Figure~\ref{fig: fig5}, high-[Mg/Fe] stars with ${\rm [Fe/H]}\geq-0.6$ display a quadrupole pattern -- a feature typically associated with bar-following stars -- similar in morphology to that of the low-[Mg/Fe] stars. The pattern weakens and becomes more centrally concentrated at lower metallicities, and is no longer detectable in the APOGEE sample for ${\rm [Fe/H]}<-0.6$. However, \cite{Liao2024insights} found that Gaia XP stars display a butterfly pattern, showing only two poles of the quadrupole pattern in their mean radial velocity field, down to a metallicity of ${\rm [Fe/H]_{XP}}\sim-1$, and that this pattern fades only in even more metal-poor stars. 

Using isolated $N$-body simulations of Milky Way-like galaxies that develop a bar through disk instability, \citet{Liao2024insights} confirmed that an initially dispersion-dominated spheroidal structure -- a pre-existing classical bulge -- can be spun up by the rotating bar to develop cylindrical rotation \citep[in agreement with][]{Saha_2012} but found that it does not exhibit the quadrupole pattern in the mean radial velocity field, suggesting that the quadrupole pattern is a distinctive feature of a bar-driven bulge.

Alltogether, both the rotational properties and the mean radial velocity fields of the observed stars suggest that the metal-intermediate $({\rm [Fe/H]}\geq-0.6)$ high-[Mg/Fe] stars prominently participate in the bar, while the transition to non bar-like properties occurs only at lower metallicities. These model-independent results are supported by more detailed orbit analysis in realistic Galactic barred potentials. Figure~\ref{fig: fig9} showed that across the metallicity range for both low- and high-[Mg/Fe] stars, ${\rm x_1}$ family tree and cloud A orbits are the primary drivers for the mean stellar rotation in the $1.5\leq R/{\rm kpc}\leq3.5$ region. Figure~\ref{fig: figA5} further confirms that these two orbit families exhibit similar angular velocity distributions in both APOGEE samples, consistent with the expected bar pattern speed. 

In the orbital density distributions (Figure~\ref{fig: fig10}), high-[Mg/Fe] stars show a transition from a peanut-shaped bulge for [Fe/H] $>-0.6$, to a boxy-bulge for $-1<{\rm [Fe/H]}<-0.6$, and finally to a spheroidal shaped bulge for PIGS stars with [Fe/H] $< -1$ \citep[see also][]{Khoperskov2024III}. Low-[Mg/Fe] stars display an even stronger peanut shape (see Figure~\ref{fig: figA1}). These results are consistent with previous studies investigating the mapping of composite disks to the bulge through bar instabilities \citep[][]{DiMatteo2016discOrigin, Debattista2017fractionation, Fragkoudi2017, Fragkoudi2018disc}. 

With $N$-body simulations and a high-resolution simulation in which all stars formed out of gas, \cite{Debattista2017fractionation} showed that stellar populations differing in their initial (before bar formation) radial velocity dispersion are separated by the bar via a mechanism they termed {\it kinematic fractionation}. The initially hotter population is lifted by the bar to larger heights, ending as a boxy bulge and forming a weaker bar, whereas the initially colder population is mapped to a stronger bar and peanut bulge. As a result, the peanut shape is generally better traced by the initially colder metal-rich stars, while the metal-poor stars trace a boxy structure \citep[see also][]{Debattista2019formation, Fragkoudi2020chemodynamics}.

In summary, the majority of high-[Mg/Fe] stars, those with ${\rm [Fe/H]}\geq-0.6$, show kinematic properties similar to the bar-driven low-[Mg/Fe] stars. The similarity is also seen in the orbital dynamics of the low- and high-[Mg/Fe] stellar samples. In the orbital density distributions, these metal-intermediate $({\rm [Fe/H]}\geq-0.6)$ high-[Mg/Fe] stars show a peanut-shaped bulge. As the metallicity decreases further, a gradual transition from bar-like to non-bar-like kinematics and morphological properties takes place. Therefore, we argue that the large majority of the high-[Mg/Fe] bulge and bar originated via secular evolution from a pre-existing high-$\alpha$ disc. Based on the chemical properties ([Fe/H], [Mg/Fe]), we associate this component with the present-day Galactic high-$\alpha$ thick disc.

\subsection{Rotational velocity at \texorpdfstring{${\rm [Fe/H]}\lesssim-1$}{[Fe/H] < -1}}\label{sec: sec4_2}
Recent studies of stars in the solar-neighbourhood have argued that stars with ${\rm [Fe/H]}\lesssim-1$ are mainly part of the proto-Galactic population, showing small but positive rotation up to a few tens of ${\rm km\,s^{-1}}$ \citep{Belokurov2022Dawn, Conroy_2022, Chandra_2024}. The net spin of the proto-Galactic population in the observations is consistent with the predictions from simulations \citep{Chandra_2024, McCluskey_2024}. 

Similarly, in the inner Galaxy ($R\leq5$ kpc), the metal-poor stars (${\rm [Fe/H]}\lesssim-1$) display significant rotational velocity \citep{Wegg_2019_RRL, Kunder2020bulgeRRL, Rix2022poor}. The origin of the rotational velocity of the metal-poor stars is still not understood. More recently, \cite{Arentsen2024orbital} found that the mean rotational velocity of PIGS stars in the Galactic bulge decreases from $\sim80\,{\rm km\,s^{-1}}$ at metallicity [Fe/H] $\sim-1$ to $\sim40\,{\rm km\,s^{-1}}$ at metallicity [Fe/H] $\sim-2.1$, and then remains nearly constant until metallicity [Fe/H] $\lesssim-2.6$\footnote{These values differ slightly from those seen in Figure~\ref{fig: fig9}, but note the different definition of apocentre distance and the inclusion of the PIGS horizontal branch stars.}. These authors associated these stars with two spheroidal populations: a more metal-rich faster rotating component and a more metal-poor slower/non-rotating component. 

Figure~\ref{fig: fig9} showed that stars associated with the ${\rm cloud\,A}$ orbits contribute most to the overall rotational velocity throughout the PIGS metallicity range. These orbits have $Z_{\rm max}/R_{\rm apocentre}\leq0.5$ (see Figure~\ref{fig: figA3}), so it is unlikely that they are part of a spheroid population. In contrast, stars associated with the ${\rm x_4}$ orbit family -- mostly spherical orbits $(Z_{\rm max}/R_{\rm apocentre}\sim1)$ -- are the most numerous over the PIGS metallicity range (Figure \ref{fig: fig7}), but contribute insignificantly ($\overline{V}_{\phi}\sim0$) to the observed overall rotational velocity. This results from a combination of spatial selection effects and the relatively large distance errors for the PIGS stars (see Section~\ref{sec: Sec_3_5_1} and Figure~\ref{fig: figA6}). By coincidence, the overall rotational velocity with the orbit-based velocity distributions is nearly the same as for the observed stars, because also the cloud A orbits are affected by the spatial selection: the contribution of the ${\rm x_4}$ orbits is then retrograde $(\sim-25\,{\rm km\,s^{-1}})$ in the mean, but this is offset by an increased prograde contribution from the cloud A orbits.

Based on these results, the two spheroidal populations model proposed by \cite{Arentsen2024orbital} to explain the trend of mean rotational velocity as a function of metallicity is uncertain. A continuous trend with metallicity seems also possible. More data are needed to investigate the origin of the net rotation in stars with [Fe/H] $\lesssim-1$, especially close to the plane, where cloud A orbits are expected to be relatively more frequent.

However, as discussed in section~\ref{sec: 41}, during the formation and evolution of bars, a stellar population that initially belonged to a non-rotating spheroidal distribution can acquire angular momentum from the bar \citep[][]{Saha_2012, Perez2017stellar, Liao2024insights}. \cite{Perez2017stellar} constructed a particle model for the spin-up by the bar of a pre-existing, centrally concentrated and non-rotating spheroid, in the context of Milky Way RR Lyrae stars. They found that a certain fraction of these stars can be trapped by the bar, thereby acquiring angular momentum, and leading to overall mild rotation of the spheroid stars.

\subsection{"Spin-up" in the bar region}
Recently, \cite{Belokurov2022Dawn} used APOGEE DR17 data to show that the mean rotational velocity of in-situ stars in the solar neighbourhood exhibit a sharp increase in the metallicity range $-1.3<{\rm [Fe/H]}<-0.9$, while the corresponding velocity dispersion decreases. The transition from an early dispersion-dominated phase to the rotation-dominated phase is referred to as the {\it Galactic spin-up} \citep[see also][]{Conroy_2022, Chandra_2024}. 

Using XP spectra from Gaia Data Release 3, \cite{Rix2022poor} reported that the mean rotational velocity for stars with cylindrical apocentre distances $3<R_{\rm apocentre}/{\rm kpc}<7$ rises steeply in the same metallicity range, similar to the Galactic spin-up described by \cite{Belokurov2022Dawn}. However, due to high dust-extinction and crowding, the optical bands of the Gaia survey data miss much of the Galactic bulge.

Figure~\ref{fig: fig9} presented the mean observed rotational velocity traced by APOGEE and PIGS stars in the region $1.5 < R/{\rm kpc} < 3.5$, including only stars with $R_{\rm apocentre}\leq5$ kpc. Also in this region of the Milky Way, the mean rotational velocity rises steeply as the metallicity increases from [Fe/H] $\sim-1.3$, a result that is driven by the stars outside the bulge. In parallel, the fraction of orbit families associated with a prograde sense of rotation (${\rm x_1}$ and ${\rm cloud\,A}$) increases rapidly while the fraction of the orbit family associated with a retrograde sense of rotation (${\rm x_4}$) decreases sharply; see Figure~\ref{fig: fig7}. Remarkably, the metallicity threshold scale of the rise matches with that of the Galactic spin-up in the solar neighbourhood, as inferred by \cite{Belokurov2022Dawn}. 

The Galactic spin-up in the solar neighbourhood marks the emergence of the present thick disc (high-$\alpha$ disc). The fact that the spin-up in the Galactic bar begins at nearly the same metallicity as in the solar neighbourhood, further supports the idea that the high-[Mg/Fe] bar component originated from the pre-existing high-[Mg/Fe] disc.

\section{Summary}
\label{sec: Summary}
Using APOGEE DR17 and the PIGS catalogue of metal-poor stars, this study examines the metallicity dependence of the dynamical properties of high-[Mg/Fe] stars within the Galactic bar region. Data from these surveys are complemented by astrometry from Gaia EDR3 and distances from the AstroNN and \texttt{STARHORSE} catalogues. For the APOGEE sample, the photometric selection function is corrected following \cite{Bovy2014Apogee} to address the first-order spatial and metallicity bias. Low- and high-[Mg/Fe] APOGEE sub-samples are defined in the [Fe/H]-[Mg/Fe] plane. To supplement the model-independent dynamical properties of these observed stars, their orbits are integrated in realistic gravitational potentials with a rotating bar from the suite of Made-to-Measure dynamical models developed by \cite{Portail_2017}. We find that our conclusions do not change across the potential models we investigated. To mitigate contamination from interlopers and halo stars, stars with cylindrical apocentre distance $R_{\rm apocentre}>5$ kpc are excluded. 

The main findings of this work are summarized as follows:
\begin{enumerate}
    \item The radial profile of the mean rotational velocity, $\overline{V}_{\phi}(R)$, for both low- and high-[Mg/Fe] stars exhibits a cylindrical rotation pattern across the bulge and the outer bar. For metal-intermediate ([Fe/H] $>-0.6$) high-[Mg/Fe] stars, $\overline{V}_{\phi}(R)$ is consistent within uncertainties with that of the bar-driven low-[Mg/Fe] stars in the bulge and only $10-20\%$ lower in the outer bar region (Figure~\ref{fig: fig3}). Also, the $\overline{V}_{\phi}(R)$ profile for intermediate-[Mg/Fe] stars matches that of low- and high-[Mg/Fe] stars in the bulge and lies between them in the outer bar. 
    
    Additionally, the $V_{\phi}-R$ distribution of high-[Mg/Fe] stars with [Fe/H] $>-0.6$ resembles that of the low-[Mg/Fe] stars with mildly increased dispersion especially in the outer bar,  and transitions to a fully dispersion-dominated distribution at low metallicity, [Fe/H] $\sim-1$ (Figure~\ref{fig: fig4}).

    \item The mean radial velocity field of high-[Mg/Fe] stars with [Fe/H] $>-0.6$ exhibits a quadrupole pattern -- a characteristic signature of a bar -- similar in morphology to low-[Mg/Fe] stars. The quadrupole pattern weakens and becomes more centrally concentrated at lower metallicities, and becomes unclear for stars with ${\rm [Fe/H]}<-0.6$ in the APOGEE sample (Figure~\ref{fig: fig5}).

    \item The orbital density distributions for high-[Mg/Fe] stars with [Fe/H] $>-0.6$ show a thick barred structure with a peanut-shaped bulge. As the metallicity decreases to [Fe/H] $\sim-1$, the bar weakens and the bulge becomes boxy. Below metallicity [Fe/H] $\sim-1$, the orbital density distribution becomes increasingly spheroidal (Figure~\ref{fig: fig10}).

    \item The fraction of orbits in the ${\rm x_1}$ family tree -- the main bar-supporting orbits -- gradually decreases from $\sim30\%$ to $\sim20\%$ across the metallicity range between the low- and high-[Mg/Fe] peaks in the [Fe/H]-[Mg/Fe] plane. Only below metallicity [Fe/H] $\sim-1$ does their fraction drop to less than $10\%$. In addition, as the metallicity increases from low-metallicity [Fe/H] $\sim-1.3$, the fraction of prograde ${\rm x_1}$ family tree and ${\rm cloud\,A}$ orbits rises sharply, whereas the fraction of retrograde ${\rm x_4}$ orbits declines significantly (Figure~\ref{fig: fig7}).

    \item The mean rotational velocity for stars in the region $1.5\leq R/{\rm kpc}\leq3.5$ increases steeply as metallicity rises from [Fe/H] $\sim-1.3$ to solar values. This result is driven by the stars outside the bulge. The overall fast rotation for [Fe/H] $>-0.9$ (APOGEE stars) is primarily driven by the $x_1$ family tree and ${\rm cloud\,A}$ orbits; while for [Fe/H] $<-1.3$, ${\rm cloud\,A}$ and retrograde ${\rm x_4}$ orbits lead to slow rotational velocity  (Figure~\ref{fig: fig9}, and section~{\ref{sec: sec4_2}}). 

\end{enumerate}

Overall, the model-independent dynamical properties of high-[Mg/Fe] stars at intermediate metallicity ([Fe/H] $>-0.6$) in the Galactic bar region  are closely similar to the low-[Mg/Fe] stars. This similarity extends to their detailed orbital dynamics. Only at lower metallicities ($-1.0 <$ [Fe/H] $<-0.6$), high-[Mg/Fe] stars show a gradual transition from bar-like to non-bar-like dynamics.

Therefore, we argue that the large majority of the high-[Mg/Fe] bulge and bar originated via secular evolution from a pre-existing high-$\alpha$ disc. Based on the chemical properties ([Fe/H], [Mg/Fe]), we associate this component with the present-day Galactic high-$\alpha$ thick disc. Comparing with a set of $N$-body models, we found that these stars rotate too rapidly to be explained by a pre-existing bulge spun up by the Galactic bar (section~\ref{sec: 41}).

The rapid increase of $\overline{V}_{\phi}\left({\rm [Fe/H]}\right)$ in the bar region for metallicity {$1.3<$ \rm [Fe/H]} mirrors the Galactic spin-up at larger radii \citep{Belokurov2022Dawn, Rix2022poor}, suggesting that the more metal-poor stars $({\rm [Fe/H]})\lesssim-1$ primarily belong to the proto-galactic population. 

\section*{Acknowledgements}  
We thank \cite{Arentsen2024orbital} for making the PIGS catalogue publicly available and for useful discussions. We thank the anonymous referee for helpful comments. AP thanks Jonathan P. Clarke for providing his script for simplifying the orbit integrations in the Made-to-Measure potentials. 

\noindent Funded by the Deutsche Forschungsgemeinschaft (DFG, German Research Foundation) under Germany's Excellence Strategy - EXC-2094 - 390783311; Excellence Cluster ORIGINS, Boltzmannstr. 2, D-85748 Garching, Germany

\noindent Funding for the Sloan Digital Sky Survey IV has been provided by the Alfred P. Sloan Foundation, the U.S. Department of Energy Office of Science, and the Participating Institutions. SDSS acknowledges support and resources from the center for High-Performance Computing at the University of Utah. The SDSS web site is \url{https://www.sdss4.org}. 

\noindent This work has also made use of data from the European Space Agency (ESA) mission Gaia (\url{https://www.cosmos.esa.int/gaia}), processed by the Gaia Data Processing and Analysis Consortium (DPAC, \url{https://www.cosmos.esa.int/web/gaia/dpac/consortium}). Funding for DPAC has been provided by national institutions, in particular the institutions participating in the Gaia Multilateral Agreement.

\noindent 

\section*{Data Availability}
The data underlying this paper will be shared on reasonable request to the corresponding author.

\bibliographystyle{mnras}
\bibliography{references}

\appendix

\section{Potential models and orbital structures}
\begin{figure*}
    \centering
    \includegraphics[scale=1.3]{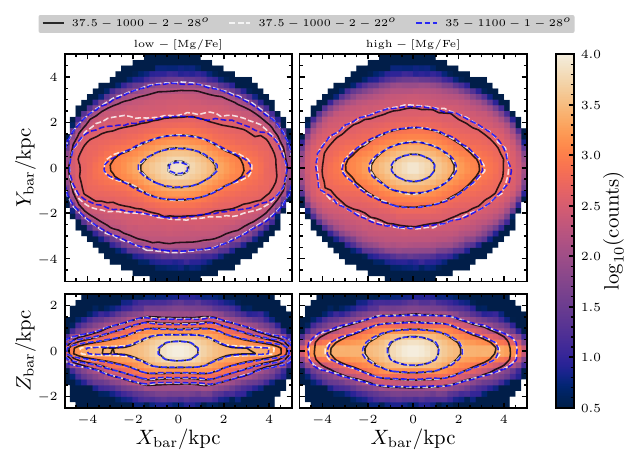}
    \caption{The orbital density distributions in the bar frame for low- and high-[Mg/Fe] (${\rm [Fe/H]>-1}$) APOGEE stars in the potential models described in section~\ref{sec: potentialOrbits}. The grey bar at the top gives the main parameters for these models: bar pattern speed (${\rm km\,s^{-1}\,kpc^{-1}}$) -- mass to clump ratio -- central in-plane mass ($\times10^{10}\,M_{\odot}$) -- bar inclination (in degrees). The colour bar and black-solid contour lines show our standard potential model, the other two potentials are shown with the white and blue dashed contour lines.  Iso-density contours range from 2.2 to 3.8 in steps of 0.4. {\bf First column:} Face-on (top row) and edge-on (bottom row) views for low-[Mg/Fe] APOGEE stars. {\bf Second column:} For high-[Mg/Fe] APOGEE stars. Orbit points with $|Z_{\rm bar}| < 1$ kpc for the face-on view, and $|Y_{\rm bar}| < 1$ kpc for the edge-on view are included. APOGEE stars are weighted by their photometric selection weights.}
    \label{fig: figA1}
\end{figure*}

In the Appendix we include additional figures that clarify specific aspects of the discussion in the main text.

Figure~\ref{fig: figA1} compares the face-on and edge-on orbital density distributions for low- and high-[Mg/Fe] $\left({\rm [Fe/H]>-1}\right)$ APOGEE stars in the three potential models described in section~\ref{sec: potentialOrbits}. For both stellar samples, the orbital densities in the three potentials are  morphologically consistent with each other. For the low-[Mg/Fe] stars, the orbital density distributions disagree slightly only in the long bar region ($|X_{\rm bar}|\geq2.5$ kpc); this difference is even smaller for high-[Mg/Fe] stars. This shows that orbital properties across the three potentials are similar.

\begin{figure*}
    \centering
    \includegraphics[scale=1.6]{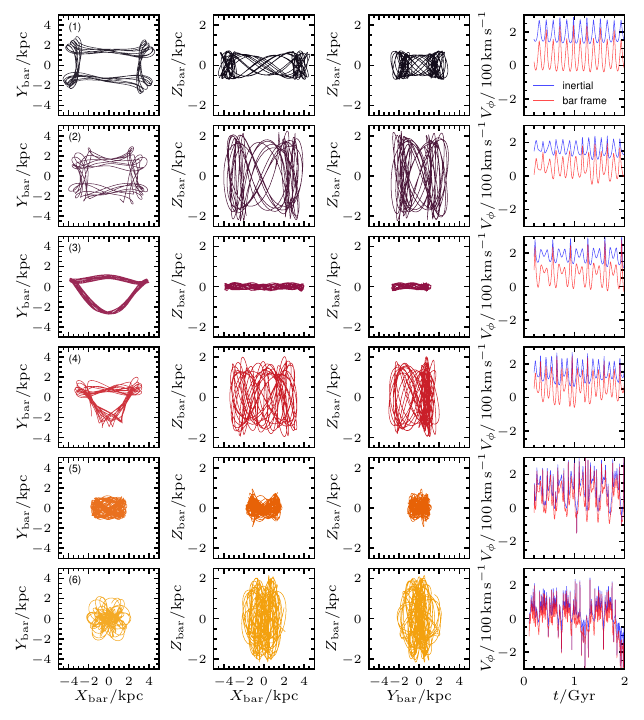}
    \caption{Each row presents the orbital density and the rotational velocity time series of a typical ${\rm cloud\,A}$ orbit, selected based on frequency ratios as shown in Figure~\ref{fig: fig6}, using triangular markers there matching the colour of the orbits shown here. Orbits (1) and (2) belong to the $f_R:f_{\phi}=4:1$ resonance sub-family of the ${\rm cloud\,A}$ family, (3) and (4) are from the $f_R:f_{\phi}=3:1$ sub-family, (5) from the ${\rm vILR},\,f_{\phi}/f_z=0.5$ sub-family, and (6) shows an orbit with higher frequency ratios.}
    \label{fig: figA2}
\end{figure*}

Figure~\ref{fig: figA2} illustrates the orbital structures and rotational velocities (both in the Galactocentric and rotating bar frames) for typical orbits in the ${\rm cloud\,A}$ family, as indicated in Figure~\ref{fig: fig6}. For the 4:1 and 3:1 resonant families, two representative orbits are shown -- one with $f_R/f_z<1$ (orbits 1 and 3) and another with $f_R/f_z>1$ (2 and 4). For both resonant families, the former are significantly thinner than the latter. The density distribution of orbit (5) -- from ${\rm vILR}$ resonant family -- is boxy shaped in the face-on view and supports the peanut structure. In contrast, orbit (6) has a spheroidal distribution. All these orbits except (6) exhibit mean prograde rotational velocities in both the Galactocentric and bar frames (right panels).

\begin{figure*}
    \centering
    \includegraphics[scale=1.2]{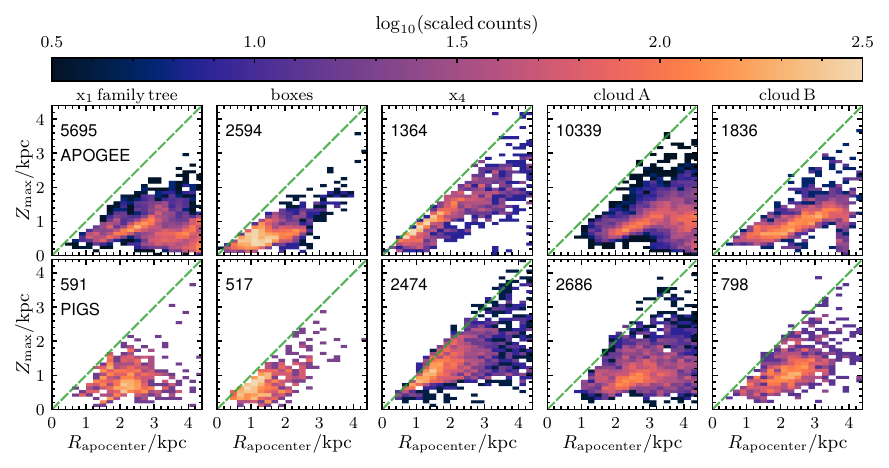}
    \caption{Distribution of APOGEE (top) and PIGS stars (bottom) associated with different orbit families, in the $R_{\rm apocentre}-Z_{\rm max}$ plane. Each panel is scaled to maintain a constant total count across all pixels. The actual numbers of stars are given in the top-left of each panel. The dashed green line represents the one-to-one relation -- the farther a star lies below this line, the more flattened its orbit structure. APOGEE stars are weighted by their photometric selection weights.}
    \label{fig: figA3}
\end{figure*}

Figure~\ref{fig: figA3} shows the distribution of APOGEE and PIGS stars grouped by their orbit families in the plane $R_{\rm apocentre}-Z_{\rm max} $. This figure highlights that ${\rm x_1}$ and ${\rm cloud\,A}$ orbits are more flattened and extend radially to the outer bar, whereas boxes and ${\rm x_4}$ orbits are more spheroidal and largely confined to the bulge. The distinctive flattened wedge-shaped distributions seen in the bar-participating ${\rm x_1}$ family tree and ${\rm cloud\,A}$ orbit families is reminiscent of the structures observed for disc stars in the solar-neighbourhood \citep[see, for instance,][]{Haywood2018disguise, Amarante2020tale}.

\begin{figure*}
    \centering
    \includegraphics[width=\textwidth]{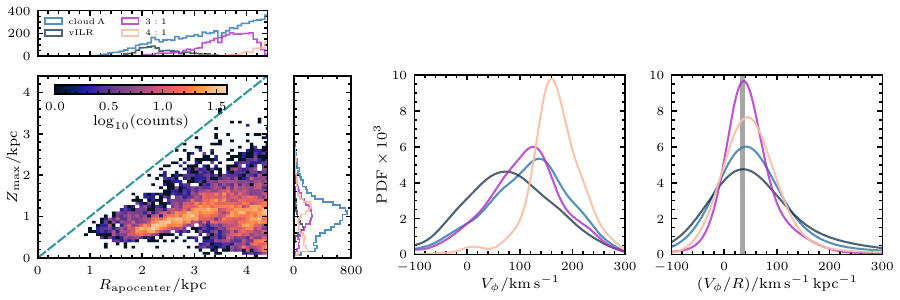}
    \caption{{\bf First column} The distribution of APOGEE stars belonging to the $\rm cloud\,A$ family in the $R_{\rm apocentre}-Z_{\rm max}$ plane along with the marginalised distributions. Marginalised distributions for near-resonant orbit subgroups are also provided, as indicated. {\bf Second and third columns}: The Galactocentric rotational velocity and corresponding angular velocity distributions of stars from the $\rm cloud\,A$ family and the near-resonant sub-groups as shown in the first column. The grey stripe in the third column indicates a range of $\left(30-40\,{\rm km\,s^{-1}\,kpc^{-1}}\right)$ for the pattern speed of the Milky Way bar. To select the resonant $\rm clould\,A$ sub-groups, we require $\rm cloud\,A$ orbits (defined in Table \ref{tab: table1}) to satisfy $\left(\frac{f_{\phi}}{f_z}=0.5\pm0.025\right)$ for $\rm vILR$, $\left(\frac{f_{R}}{f_{\phi}}=3\pm0.25\right)$ and $\left(\frac{f_{R}}{f_{\phi}}=4\pm0.25\right)$ for 3:1 and 4:1 families, respectively. These ranges for the frequency ratios are chosen for better statistics; the orbit-structural and rotational properties remain unchanged for slightly larger or smaller ranges. APOGEE stars are weighted by their photometric selection weights.}
    \label{fig: figA4}
\end{figure*}

Figure~\ref{fig: figA4} is similar to Figure~\ref{fig: figA3} but focuses on $\rm cloud\,A$ and its near-resonant (3:1, 4:1, and ${\rm vILR}$)  orbits in the APOGEE sample, showing their distributions in the $R_{\rm apocentre}-Z_{\rm max}$ plane (first column), as well as their Galactocentric rotational and angular velocity distributions (second and third columns, respectively). This figure shows that while stars near the $\rm vILR$ resonance remain confined to the bulge, those near the 3:1 and 4:1 resonances extend radially to the outer bar. Moreover, stars near the $\rm vILR$ resonance exhibit lower rotational velocities and larger dispersion compared to those near the 3:1 and 4:1 resonances. Nonetheless, the peaks of the angular velocity distributions for all stars belonging to the $\rm cloud\,A$ family and its near-resonant sub-groups are nearly aligned and consistent with the expected range of the pattern speed for the Milky Way bar.

\begin{figure*}
    \centering
    \includegraphics[scale=1.55]{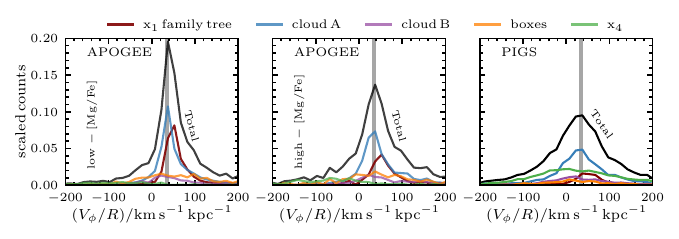}
    \caption{The Galactocentric angular velocity $\left(V_{\phi}/R\right)$ distributions for APOGEE low- and high-[Mg/Fe] stars (left and middle panels, respectively) and PIGS (right) stars associated with different orbit families. The coloured curves show the contribution of the orbit families in bins of angular velocity which add-up linearly to the total (black line). The grey stripes indicate the Milky Way's bar pattern speed range $30-40\,{\rm km\,s^{-1}\,kpc^{-1}}$. APOGEE stars are weighted by their photometric selection weights.}
    \label{fig: figA5}
\end{figure*}

Figure~\ref{fig: figA5} shows the Galactocentric angular velocity ($V_{\phi}/R$) distributions of low- and high-[Mg/Fe] APOGEE stars (left and middle) and PIGS stars (right), grouped by orbit family. In both APOGEE subsamples, the ${\rm x_1}$ family tree and cloud A orbits dominate in number. Their angular velocity distributions peak near the expected bar pattern speed range, slightly higher for the ${\rm x_1}$ due to stronger streaming motions. Sub-dominant families (boxes, cloud B, ${\rm x_4}$) exhibit broader distributions. In contrast, PIGS stars are mainly associated with the cloud A and retrograde ${\rm x_4}$ orbit families; while those in the cloud A family still peaks near the pattern speed but their distribution is markedly broader than in APOGEE. The ${\rm x_4}$ stars in PIGS has a retrograde peak with even broader distribution. This suggests similar orbital dynamics between low- and high-[Mg/Fe] stars, but notable differences from the more metal-poor PIGS population (${\rm [Fe/H]}\lesssim-1$).

\begin{figure*}
    \centering
    \includegraphics[]{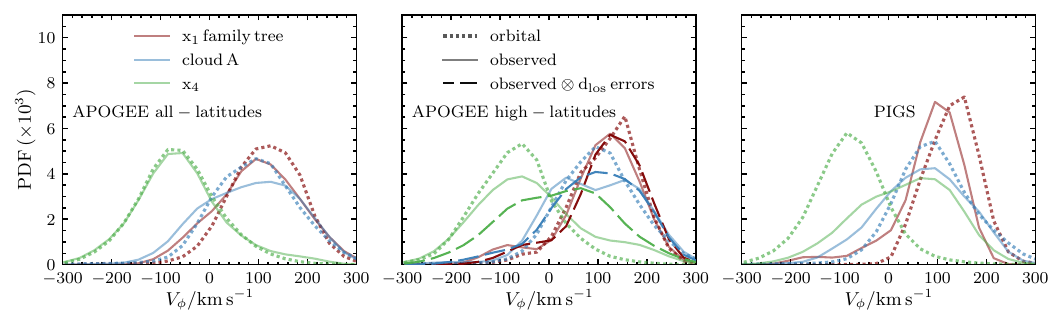}
    \caption{The observed (solid) and orbit-based (dotted) Galactocentric rotational velocity distributions in the range of metallicity $\left(-1.3 \leq {\rm [Fe/H]} \leq -0.9\right)$ and line-of-sight distance $(6\leq d_{\rm los}/{\rm kpc}\leq9)$, common to both surveys. The left panel shows the distributions for all APOGEE stars with these cuts, the middle panel for only those with absolute latitude $(7^{\circ} < |b| < 12^{\circ})$, and the right panel for PIGS stars in the same latitude range. The dashed curves in the middle panel show APOGEE stars convolved with a Gaussian distance error distribution (${\rm mean}=15\%$, ${\rm sigma}=5\%$) corresponding to the larger distance errors in PIGS. In all panels, stars with $R_{\rm apocentre} > 5$ kpc are excluded. APOGEE stars are weighted by their photometric selection weights.}
    \label{fig: figA6}
\end{figure*}
Figure~\ref{fig: figA6} compares the observed and orbit-based rotational velocity distributions for the APOGEE and PIGS stars in the range of metallicity $-1.3\leq{\rm [Fe/H]}\leq-0.9$, latitude $7^{\circ}\leq|b|\leq12^{\circ}$ and line-of-sight distance $6\leq d_{\rm los}\,/\,{\rm kpc}\leq9$ common to both surveys, to allow a more direct comparison. For visual clarity, only the three dominant orbit families in this metallicity range are shown. These selections make the resulting observed and orbit-based rotational velocity distributions in APOGEE (middle panel) more consistent with their PIGS counterparts (right panel) across the orbit families. Notably, the observed distribution for APOGEE ${\rm x_4}$ stars (solid green curve) develops a stronger prograde tail, though it remains retrograde in the mean. In addition to these selections, we convolved the APOGEE stars with a Gaussian distance error distribution corresponding to the larger distance errors in PIGS (dashed lines in the middle panel). Interestingly, this convolution brings the observed rotational velocity distribution for the ${\rm x_4}$ stars in APOGEE into good agreement with that of PIGS, while having minimal impact for the ${\rm x_1}$ family tree and cloud A orbit families. Since the ${\rm x_4}$ stars reside closer to the Galactic center, their rotational velocity distribution is more susceptible to distance uncertainties than for the other orbit families, because stars can change the sign of their rotational velocities when they are shifted across the Galactic center distance through distance errors. Thus, the combined effects of larger distance uncertainties and spatial selection effects of the PIGS survey explain the curious observed rotational velocity distribution of the ${\rm x_4}$ stars.

\bsp
\label{lastpage}
\end{document}